\documentclass[10pt,conference,final,comsoc]{IEEEtran}
\IEEEoverridecommandlockouts
\usepackage{cite,flushend}
\usepackage{amsmath,amsfonts}
\usepackage{amsfonts,dsfont}
\usepackage{algorithmic}
\usepackage{graphicx}
\usepackage{textcomp}
\usepackage{xcolor}
\usepackage{slashbox}
\usepackage{braket}
\usepackage{multirow}

\usepackage[T1]{fontenc} 
\usepackage{amsmath}
\usepackage[cmintegrals]{newtxmath}
\usepackage{bm} 

\newcommand{\Real}{\mathds R}

\begin{document}

\title{Optimal Parametrization of the Gale-Shapley Preallocation Method for Combinatorial Auction-based Channel Assignment
\thanks{The work of D. Csercsik is funded by the Hungarian Academy of Sciences under its Momentum Programme LP2021-2.
The work of E. Jorswieck is partly funded by the German Research Foundation (DFG) under grant JO 801/24-1 and by the Federal Ministry of Education and Research (BMBF) within the 6G Research and Innovation Center (6G-RIC) under support code 16KISK031.}
}

\author{\IEEEauthorblockN{ D\'{a}vid Csercsik}
\IEEEauthorblockA{
\textit{Institute of Economics} \\
\textit{Centre for Economic and Regional Studies}, and \\
\textit{Faculty of Information Technology and Bionics} \\
\textit{P\'{a}zm\'{a}ny P\'{e}ter Catholic University}\\ 
Budapest, Hungary\\
csercsik.david@krtk.hun-ren.hu}
\and
\IEEEauthorblockN{Eduard Jorswieck}
\IEEEauthorblockA{\textit{Institute for Communications Technology} \\
\textit{Faculty of Electrical Engineering,} \\
\textit{Information Technology, Physics} \\
\textit{TU Braunschweig}\\
Brunswick, Germany \\
 e.jorswieck@tu-bs.de}}


\maketitle


\begin{abstract}
Algorithms based on combinatorial auctions show significant potential regarding their application for channel assignment problems in multi-connectivity ultra-reliable wireless networks. However the computational effort required by such algorithms grows fast with the number of users and resources. Therefore, preallocation-based combinatorial auction represents a promising approach for these setups. The aim of the preallocation is to constrain the number of bids submitted by participants in the combinatorial auction process, thus reducing computational demands and enabling numerical feasibility of the auction problem. Reduction of bid number is achieved via limiting the number of items (channels) considered by auction participants (tenants) in their bids. Thus the aim of preallocation is to non-exclusively assign channels to tenants. This assignment serves as a basis for the later bid generation in the auction procedure.
In this paper we analyze the optimal parametrization of the many-to-many Gale-Shapley preallocation method and formulate recommendations for optimal performance. Numerical assessments illustrate that the appropriate preallocation has significant impact on the performance and computational demand.
\end{abstract}


\section{Introduction}

\subsection{Beyond 5G and URLLC}
The design of the 5G new radio (NR) wireless standard enables multi-service communications. Besides the services connected to traditional human type communications (HTC) via the enhanced mobile broadband (eMBB) service class, 5G NR has two additional service classes, which support machine type communication (MTC). These are namely the novel ultra-reliable low-latency communications (URLLC) and massive MTC (mMTC) service classes, from which URLLC is arguably the most challenging \cite{Sutton2019}.

In 5G NR, this challenge is mainly addressed by a new 5G numerology, i.e., differentiated parameter choices for the orthogonal frequency division multiplex (OFDM) access \cite{Segura2021}. As outlined in the conclusions of \cite{Segura2021}, the reliability needs further improvements which cannot be realized by the new numerology alone, but instead requires additional technologies, e.g., multi-link connectivity, which will provide higher reliability and, in certain cases, lower latency.

URLLC is regarded as a cornerstone of emerging beyond 5G and 6G mobile wireless communications systems \cite{chen2018ultra}. There are several mission-critical real-time applications, where URLLC requirements are unavoidable \cite{Hossler2018Mission}, like smart-factory/industrial automation, augmented-reality assisted surgery or real-time control of autonomous vehicles.

One potential way to enhance the reliability of the communication architecture, and thus achieve URLLC requirements, is the application of redundancy or diversity, i.e. using multiple communication paths simultaneously to connect communicating objects (\emph{tenants} in the following) to the system \cite{suer2019multi}.
Due to the limited accessibility of resources in such multi-connective, multi-user environments, the problem of channel allocation/assignment naturally arises.
In addition, the computational complexity of the respective assignment problem becomes more and more a challenge as the number of tenants and the number of communication paths is increased \cite{Manap2020}.

\subsection{Related Literature}

Besides deep learning approaches, which have been recently proposed as a potential solution for the resource allocation problem \cite{She2021}, game-theory based methods and algorithms have also been applied for such tasks in 6G \cite{Elsayed2019}. In general, game theoretic applications are well represented in the literature of wireless resource allocation problems (for a review see \cite{ALTMAN2006286} or \cite{CHARILAS20103421})

In recent years, a number of elegant and efficient channel auction algorithms are developed for dynamic channel assignments and important challenges such as strategy proofness and computational complexity are studied \cite{Wu2016}. Low-complexity matching-based channel assignment algorithms are developed to find stable matchings in cognitive radio networks \cite{Mochaourab2015,han2017matching}. A low complexity channel allocation scheme for URLLC services is proposed in \cite{Ben2021}. Two distributed spectrum auction mechanisms are proposed in \cite{Yang2019} namely distributed VCG and FAITH. Distributed VCG implements the celebrated Vickrey-Clarke-Groves mechanism in a distributed fashion to achieve optimal social welfare, at the cost of exponential communication overhead. In contrast, FAITH achieves sub-optimal social welfare with tractable computation and communication overhead. 

In channel assignment problems, channels may be either considered as indivisible goods or, they may be further divided, e.g. by time-sharing methods. Regarding the former case, recent results  \cite{csercsik2023preallocation} have shown that combinatorial auction (CA) based methods \cite{de2003combinatorial} have the potential to provide channel allocations, which are efficient and fair in the same time.
Due to computational limitations however, the CA method can not be directly applied unless the number of channels is sufficiently low.


\subsection{The Concept of Preallocation}

\begin{figure}
  \centering
  \includegraphics[width=0.8\linewidth]{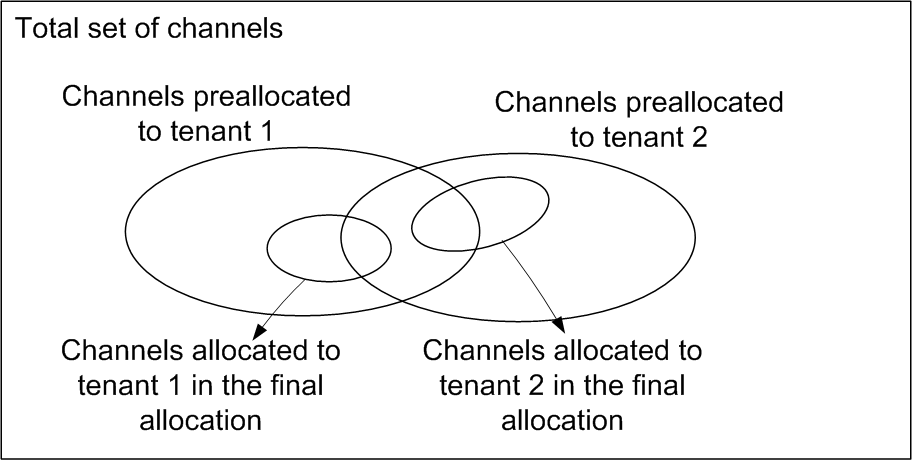}\\
  \caption{The relation of channel sets after the preallocation and the final allocation in the case of two tenants.}\label{Prealloc_concepts}
\end{figure}

The method proposed in \cite{csercsik2023preallocation} overcomes this problem by applying a \emph{preallocation} of channels. Preallocation means that before the combinatorial auction, a \emph{non-exclusive} assignment of channels to tenants is performed (i.e. a channel is potentially assigned to multiple tenants as well), and each tenant considers only the subsets of channels preallocated to it in the bidding process. In other words, the search space of the CA optimization problem is constrained. The relation of the preallocated and finally allocated channel sets is depicted in Fig. \ref{Prealloc_concepts}.

In this setup, the first critical task of the preallocation process is to limit the computational demands of the consecutive CA, by limiting the number of bids submitted. The second task is naturally to ensure an appropriate preallocation in the sense, that optimum achieved by the CA later on the limited bid set has to be efficient. 

\subsection{Motivation}

The method proposed in \cite{csercsik2023preallocation} uses the many-to-many version of the famous Gale-Shapley (GS) algorithm \cite{gale1962college}. The paper \cite{csercsik2021Heuristics} argues, that using the many-to-may version of the GS for (M2MGS) preallocation provides slightly better results compared to the more straightforward simple approaches of distance-based preallocation. However, the paper \cite{csercsik2021Heuristics} considers only the total resulting capacity of the system as a performance measure for comparison. Furthermore, it does not assume the potential blockage of channels in the simulation setup, which may easily arise, due to e.g. obstacles. In addition, this very initial comparison of these two preallocation methods does not analyze, how the proposed algorithms scale up for problems of increased size.

Furthermore, the paper \cite{csercsik2023preallocation} assumes a fixed quota for both the channels and the tenants (6) in the GS-based preallocation process. In this paper we study how the quota parameters affect the output of the preallocation process and thus the overall efficiency of the preallocation based CA process, and what guidelines may be formulated regarding their optimal choice.
The optimal choice of these parameters may be critical in practical applications, as the set of preallocated channels constraints the search space of the CA algorithm, thus significantly affects its potential optimum and its computational time as well.

\subsection{Contributions}

Our main aim in the current paper is to analyze the optimal parameterization of the M2MGS preallocation method. This corresponds to the task of optimally choosing the quotas in the many-to-many version of the GS. While in addition to the total resulting utility of tenants, there are multiple potential aspects corresponding to the efficiency of the resulting allocation (e.g. fairness, as discussed in \cite{csercsik2023preallocation}), in this paper we restrain ourselves to interpret optimality and performance exclusively in the context of the total resulting utility and required computational time. The user-centric view on the resource allocation has gained attention recently \cite{CHEN2021} in the context of cell-free massive MIMO networks. The utility function describes how the utility (i.e. satisfaction) of the actual tenant depends on its actual resulting capacity / throughput value. The utility function includes a minimal and a maximal required capacity value: If the capacity is below the minimal value, the resulting utility is zero, while if the capacity exceeds the maximal required capacity value, the utility saturates at the value of 1. Considering the minimal and maximal capacity requirement of each tenant makes the problem more challenging, and more realistic in the same time, compared to studies in which the sole measure of efficiency is the total allocated capacity of the system (as e.g. \cite{Hossler2019matching}). 

\section{Model and Problem Statement}\label{sec_model}


 The number of tenants and the number of BSs are denoted by $n_T$ and $n_{BS}$ respectively. In the considered setup, each BS serves a finite number of channels for tenants. The column vector $n_{ch} \in \Real ^{n_{BS}}$ holds the number of channels offered by BSs. The $i$-th element of $n_{ch}$, denoted by $n_{ch}^i$ equals to the number of channels offered by $BS_i$, while $n^T_{ch}$ denotes the total number of channels (i.e. $\sum_i n_{ch}^i = n^T_{ch}$). Channels in the assumed framework are considered as indivisible goods.

For the description of the assignment, we use the binary assignment matrix, denoted by $A \in \Real ^{n_T \times n^T_{ch}}$. While the rows of $A$ correspond to tenants, the columns of $A$ correspond to channels: The $k,j$-th entry equals one, i.e. $A(k,j)=1$, if channel $j$ is assigned to tenant $k$, else $A(k,j)=0$. Since in the final assignment each channel may be assigned to maximum one tenant, the sum of the columns of $A$ is less or equal to 1.

Following \cite{csercsik2023preallocation}, we use the concept of the \emph{connectivity function}. The connectivity function of tenant $k$ (denoted by $\rho_k(A)$) defines the resulting capacity of tenant $k$ in the case of the allocation described by $A$. Let us emphasize that in the current framework, we assume that the resulting capacity of any tenant depends only on the set of channels assigned to it, and it is not affected by the channels allocated to other tenants.
This assumption may be regarded as neglecting intra-cell interference, which may be plausible in the case of either orthogonal channel allocation in frequency, or time domain or, if the same time-frequency resource is assigned to two users, in the case of the application of modern spatial signal processing techniques such as spatial division multiple access (SDMA) or massive MIMO transmissions.

Thus we write the resulting formula for the capacity of tenant $k$ ($C_k$), where $A_{(k,.)}$ denotes the $k$-th row of $A$, describing the set of channels allocated to tenant $k$, as follows
\begin{equation}\label{eq_Ck}
  C_k=\rho_k(A_{(k,.)}).
\end{equation}
The explicit calculation of the connectivity function is described in \cite{csercsik2023preallocation}.
Here we restrain ourselves to underline some important properties of $\rho_k$. First of all, $\rho_k$ may be efficiently calculated, if all parameters, and the set of the channels assigned to tenant $k$ ($A_{(k,.)}$) are known. However in general it is not trivial to compute it in closed form. Second, the function $\rho_k$ is non-decreasing as the set of channels assigned to tenant $k$ is expanded.

Moreover, in the case of few channels, the function is typically supermodular. If e.g. originally one channel is assigned to the tenant, and it receives an other channel, the level of redundancy is significantly enhanced, and as a consequence the resulting capacity is largely increased. 
This is due to the fact that the framework assumes independent channels, and outage of a tenant requires all of its assigned channels to be in outage.
On the other hand, if the tenant has already multiple channels (i.e. 3 or 4), the addition of further channels does not increase the level of diversity and thus the resulting capacity very much, thus in such cases $\rho_k$ typically behaves as a submodular function. In other words, as the set of assigned channels grows, the value of $\rho_k$ tends to saturate.
To give an impression about the connectivity function, a simple numerical example is included in \cite{csercsik2021Heuristics}.

The utility of an actual tenant depends on its resulting capacity. Similarly to \cite{csercsik2023preallocation}, following the 'TCP interactive user' profile of \cite{SHI20082257}, we use the utility function described by 
\begin{align}\label{eq_utility_fncn}
U_k(x)= &\begin{cases}
\frac{\log(x/C_k^{min})}{\log(C_k^{max}/C_k^{min})} \frac{\textrm{sgn}(x-C_k^{min})+1}{2} &\text{if}~  x \leq C_k^{max}\\
1 &\text{if}~ x>C_k^{max}\\
\end{cases}.
\end{align}
The function described in e.q. (\ref{eq_utility_fncn}) holds two parameters, namely $C_k^{min}$ and $C_k^{max}$ for each tenant $k$, corresponding to the previously mentioned minimal and maximal required capacity values respectively.

Considering the above assumptions and notations, (\ref{eq_assignment_opt}) describes the considered channel assignment problem, where the aim is to maximize the total resulting utility.
\begin{align}
    &\max_{0 \leq A \leq 1} \sum_{k=1}^{n_T} U_k(\rho_k(S_k)) \nonumber \\
    & \text{s.t. } \sum_k A(k,m) \leq 1 ~~~ \forall m ~(1 \leq m \leq n_{ch})
    \label{eq_assignment_opt}
\end{align}
The problem (\ref{eq_assignment_opt}) defines an integer programming problem (or a combinatorial optimization problem, since the space of the possible solutions is finite), where the objective function is nonlinear.

\section{Combinatorial Auction and Preallocation}

The principles of combinatorial auction (CA) are described in \cite{de2003combinatorial}. The CA framework assumes that bidders (in this case tenants) submit bids for  for each possible subset of goods (channels), according to their own evaluation of bundles (in this case the resulting connectivity).
The bid announced by player (tenant) $k$ for the bundle  $S$ of channels is denoted by $b_k(S)$. In our case, this value will be determined by $U_k(\rho_k(S))$. $y(S,k)$ denotes the acceptance indicator of the bundle $S$ for the participant $k$, i.e. $y(S,k)=1$ if the bundle $S$ is assigned to player $k$, and 0 otherwise. Following the submission of bids, an integer optimization problem described by (\ref{CA_basic_form}) is performed in order to maximize the value of accepted bids.
\begin{align}
    & \text{max}\sum_{k} \sum_{S \subseteq CH}b_k(S)y(S,k) \label{CA_basic_form} \\
    & \text{~s.t.~} \sum_{S \ni m} \sum_{k} y(S,k) \leq 1~~\forall m \in CH,  ~~ \sum_{S \subseteq CH} y(S,k) \leq 1 ~\forall k  \nonumber
\end{align}

The first constraint in (\ref{CA_basic_form}) ensures that overlapping goods are never assigned, while the second constraint ensures that no bidder receives more than one bundle

It is easy to see that if all players consider all possible subsets of all channels in the bundles for they submit their bids, the solution of \ref{CA_basic_form} coincides with the solution of \ref{eq_assignment_opt}. However, even in the case of relatively low total number of channels (e.g. 15), this results in computationally infeasible problems, since the number of bundles grows exponentially with the number of channels. E.g. in the case of 15 channels, there are $2^{15}-1$ bundles for each tenant. In such cases the optimal allocation can not be determined due to computational limitations, but the efficiency of the allocation determined by the CA based method may be compared to the results of other state-of-the-art methods as described in \cite{csercsik2023preallocation}. Let us furthermore note that these bundles must be also evaluated to determine the respective bid. This practically means a channel state estimation for the corresponding wireless channels, which would imply additional time delays in practical applications. 

\subsection{Preallocation of Channels}

In this paper, we assume that the method of \emph{channel preallocation} is used to overcome the above computational limitation of the CA approach applied in the case of the channel assignment problem.

Preallocation means that we limit the number of channels for each tenant, which are considered in the bids of the actual tenant. In other words, for every tenant $k$, we define a subset of channels ($CH^k$) considered by the tenant in the auction process. In other words, the bundles, for which tenant $k$ places bids in the CA process are derived as all possible nonempty subsets of $CH^k$ (thus tenant $k$ is neglecting the rest of the channels). The task of the preallocation process is to determine these, in general non-disjoint $CH^k$ sets. In other words, the above consideration means that \emph{pre}allocation of channels is non exclusive (i.e. one channel may be preallocated to multiple tenants as well).

Regarding the assignment matrix $A$, we can use the very same structure to describe a preallocation. As discussed before, if the sum of the columns of $A$ is less or equal than 1, the assignment is univalent, i.e. each channel is assigned to maximum one tenant. 
In the process of preallocation, we relax this assumption, and
we denote the (typically non univalent) assignment matrix resulting from the preallocation process by $A^p_M$, where the lower index $M$ corresponds to the preallocation method used.

Following the preallocation procedure (see the algorithms applied for this step later), the tenants submit their bids for the relevant bundles determined by $CH^k$, and the CA described in eq. (\ref{CA_basic_form}) is solved to determine the final (univalent) allocation, described by a matrix $A$, for which the column sums are already bounded by 1.

\begin{figure}
  \centering
  \includegraphics[width=1.0\linewidth]{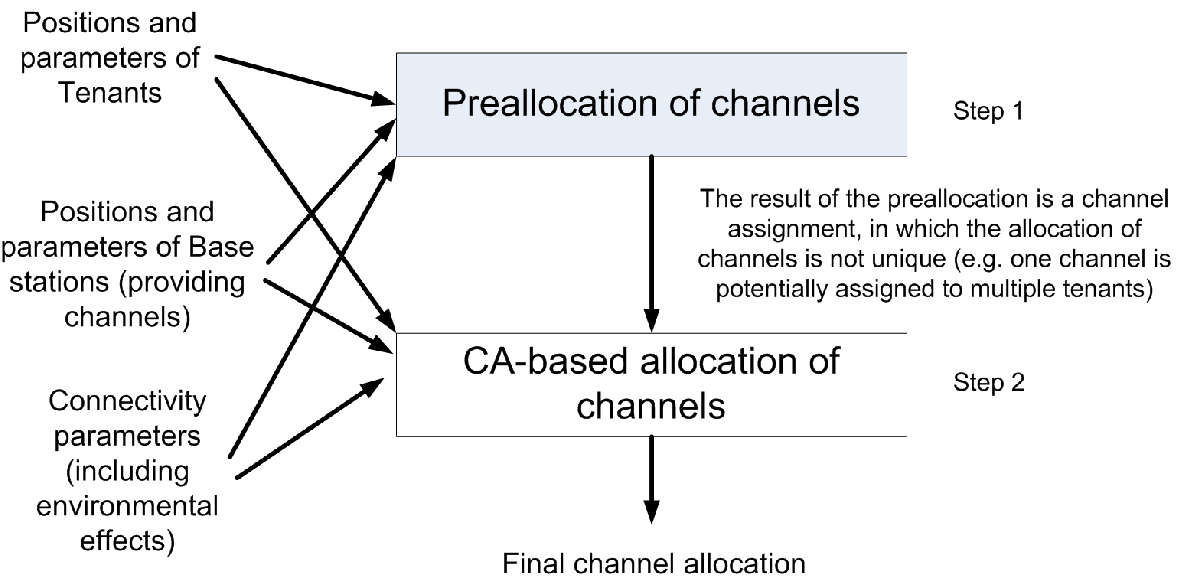}\\
  \caption{Overview of the CA-based allocation process.}\label{Fig_Allocation_process_steps}
\end{figure}

The two-step channel allocation procedure (preallocation and CA-based allocation) is illustrated in Fig. \ref{Fig_Allocation_process_steps}. 

\subsection{Preallocation via the many-to-many Gale-Shapley (M2MGS)}\label{sec_M2MGS_PA}

The preallocation of channels is carried out in the case of M2MGS as follows. In the initial step, the preferences of tenants over channels, and channels over tenants are set up. We assume that the basis of preference relations is the single-connectivity value (SCV), defined by $\rho_k(A^s_{(k,.)})$, where $A^s_{(k,.)}$ is a row vector of unit length, containing a single nonzero entry at the position corresponding to the channel in question. Tenants prefer channels providing higher SCV, and channels prefer tenants with higher SCV. Tie-breaks, which naturally emerge since BSs are serving multiple individual channels with same parameters, are resolved by randomization.

The method has two important parameters, namely the quota of tenants, denoted by $q_T$ and the quota of channels, denoted by $q_{ch}$. As we assume that the maximal number of bids submitted per tenant is bounded in the CA process, we assume $q_T \leq \bar{n}_{ch}$ ($\bar{n}_{ch}=8$ is assumed in this paper). Furthermore, as we look for \emph{pre}allocations, where the assignment of channels is potentially non-unique, we assume $1 < q_{ch}$.

After the preference lists of tenants and channels are complete, the algorithm is executed according to the principle of \cite{gale1962college}. Each channel proposes to the $q_{ch}$ most preferred tenant. Each tenant puts the $q_T$ most preferred offers on hold, and rejects the rest. Rejected channels propose to the next tenant in their preference list (until the list runs out), and so on, until no offers are rejected.


\section{Numerical Assessments}\label{sec_params}

\subsection{Simulation Setup and Parameters}

In this study, we use the model parameters and notations detailed by \cite{csercsik2023preallocation}, except for the matrix $K$, which contains the $K_{k,i}$ values. These values describe the fading of channels due to the obstacles in the simulation space. $K_{k,i}$ corresponds to the fading experienced between tenant $k$ and BS $i$.  The case $K_{k,i}=K_{ref}=14.1~dB$ corresponds to a clear line-of-sight (LOS) path, while a decreased value corresponds to blockage. In this paper we consider a moderately complex environment, in the sense that the 30\% of all $K_{k,i}$ values are decreased by 20 -- 80 \% (the rest are equal to $K_{ref}$).

We consider three setups of different sizes in the simulation.
\begin{itemize}
  \item \textbf{Small-scale (SS) setup:} The small scale setup is identical to the setup used in \cite{csercsik2023preallocation}. In this setup we assume a 100m $\times$ 50m rectangular area, with 6 tenants in random positions, and 8 BSs, located in random positions on the boundary of the area. In this setup, each BS offers 1--3 identical channels, but the total number of channels is no more than 20.
  \item \textbf{Medium-scale (MS) setup:} In this setup we assume a 120m $\times$ 70m rectangular area, with 12 tenants, and 12 BSs. In this setup, each BS offers 2--5 identical channels, but the total number of channels is no more than 45.
  \item \textbf{Large-scale (LS) setup:} In this setup we assume a 150m $\times$ 100m rectangular area, with 20 tenants, and 16 BSs. In this setup, each BS offers 3--6 identical channels, but the total number of channels is no more than 60\footnote{Note that in some applications, large-scale scenario comprises many more tenants and channels, e.g., in cellular hot-spot scenarios. However, the CA algorithm is applicable in these scenarios as well, if the number of bids is sufficiently low. E.g. the algorithm could handle 100 tenants if each submits only $2^6$ bids. }
\end{itemize}

\subsection{Numerical Results}\label{sec_results}

\begin{figure}
  \centering
  \includegraphics[width=0.48\linewidth]{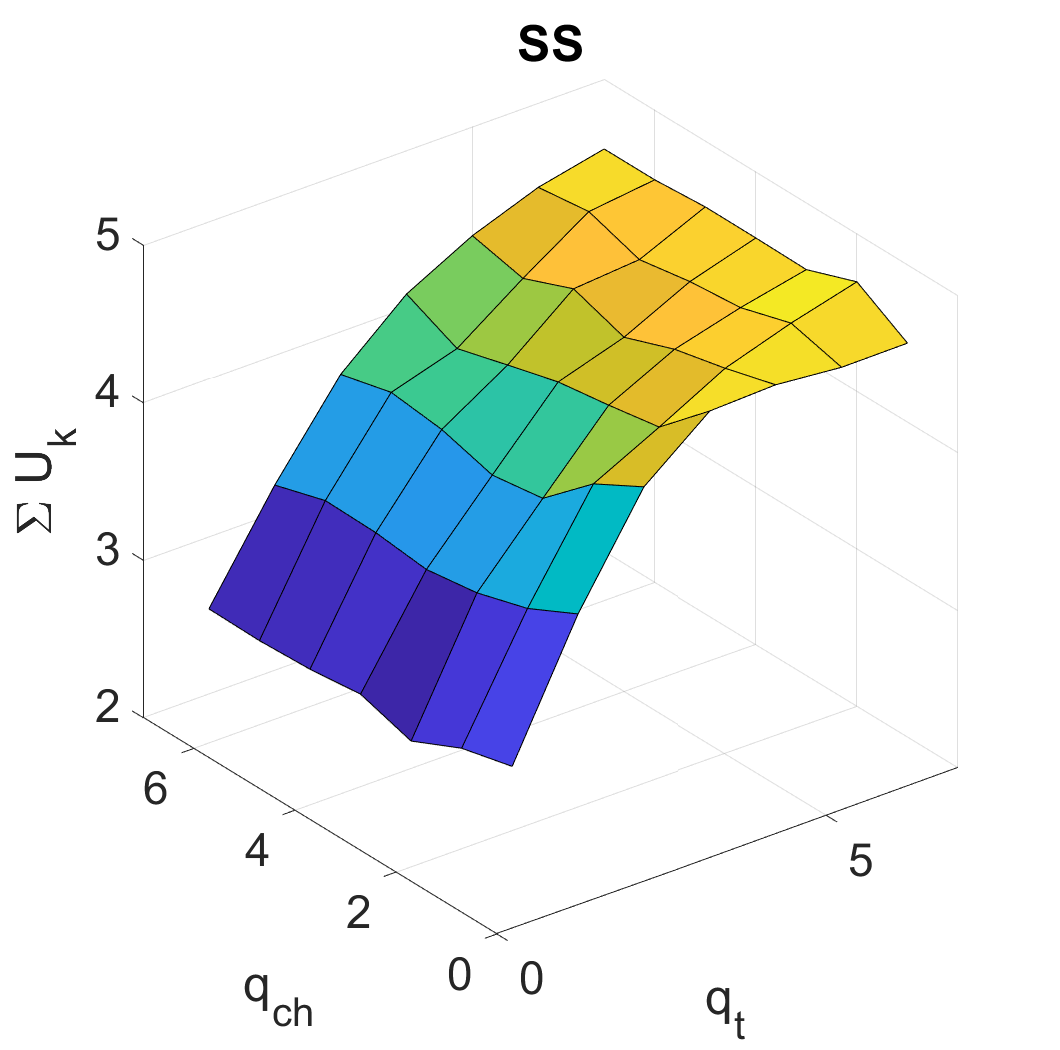}
  \includegraphics[width=0.48\linewidth]{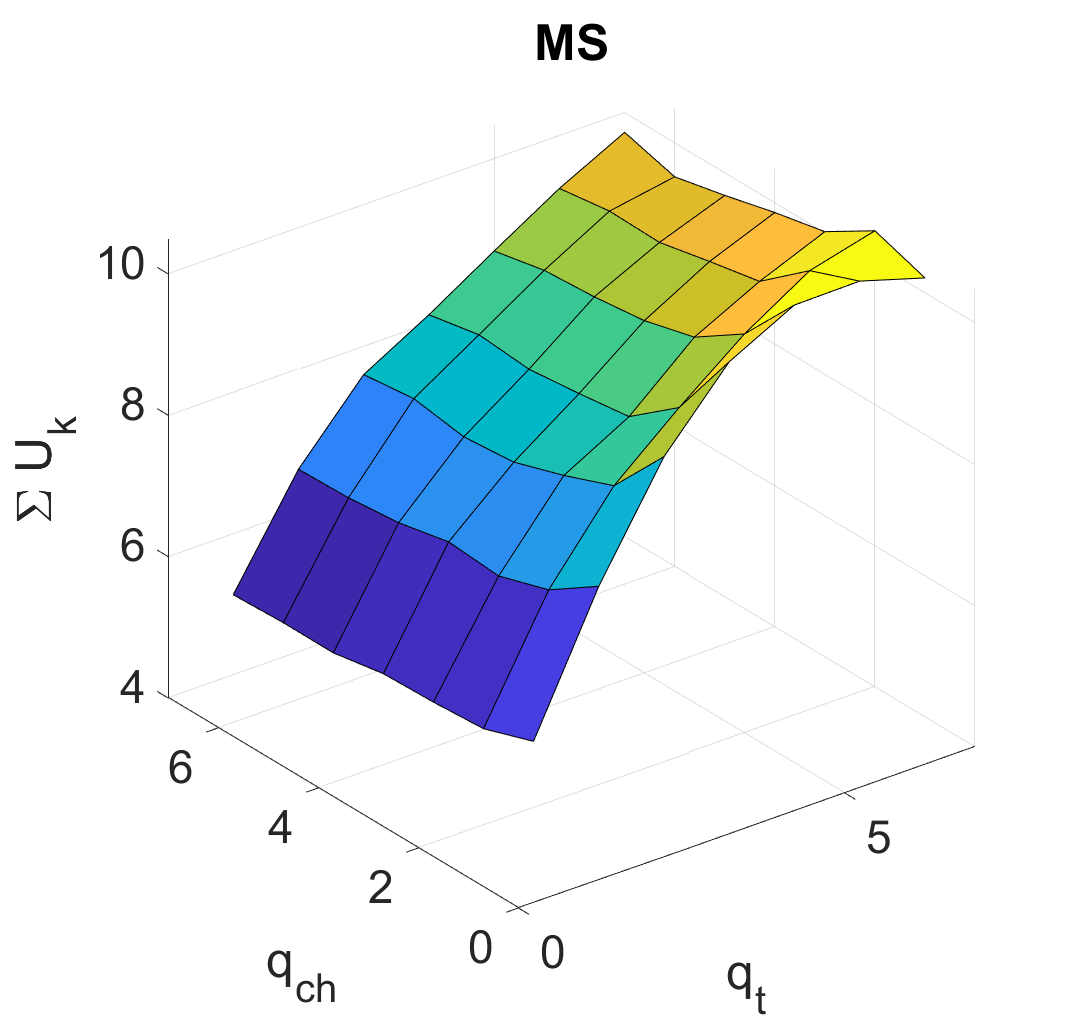}
  \\
  \includegraphics[width=0.48\linewidth]{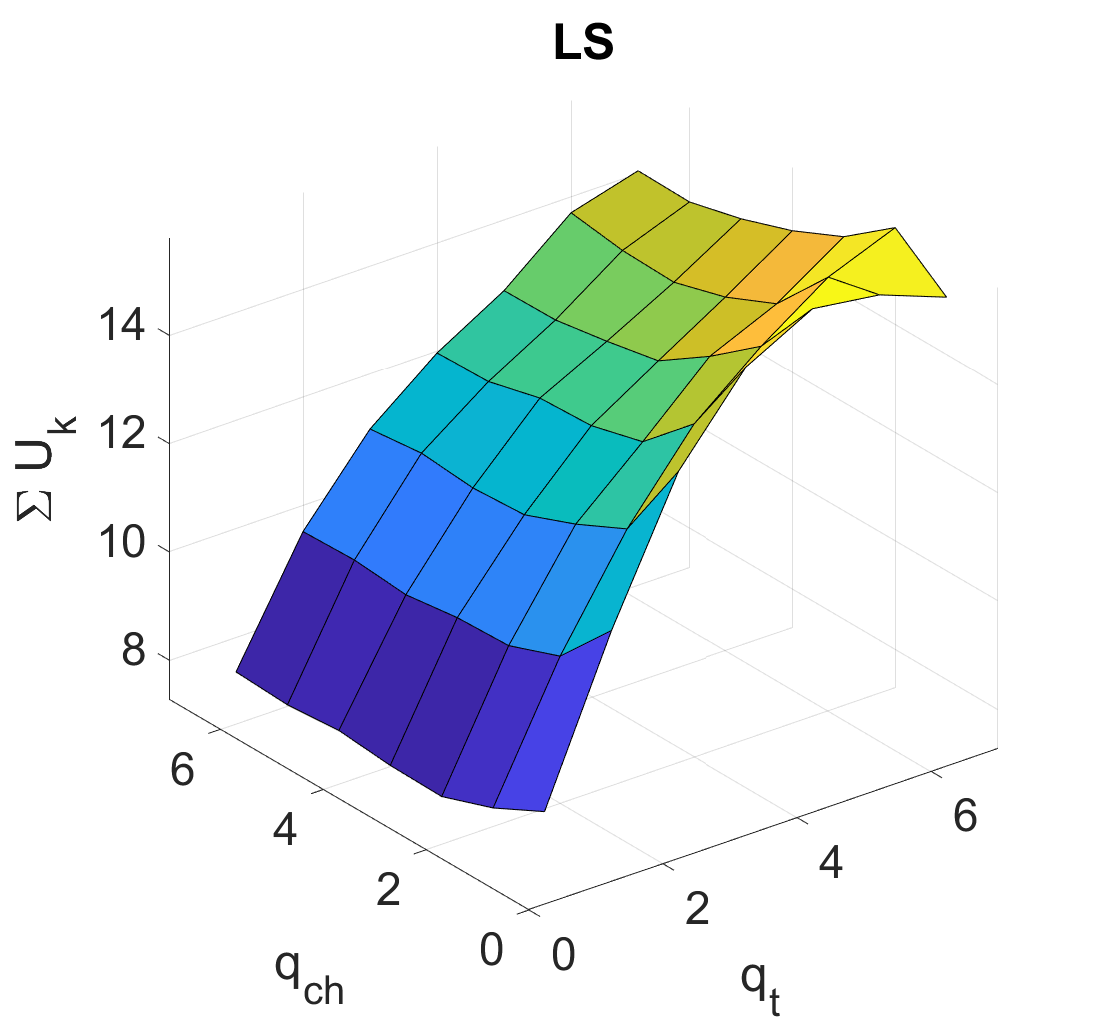}
  \caption{Resulting total utility ($\sum U_k$) values, in the case of various $q_T$ and $q_{ch}$ values. Average values of 200 simulation runs.}\label{Fig_U_k}
\end{figure}

The most straightforward measure of any preallocation method is whether it serves as a good search space for the CA in the second step of the allocation process. According to this, let us first analyze, how the choice of $q_T$ and $q_{ch}$ affects the resulting total utility, if the 'standard CA' method described by \cite{csercsik2023preallocation} is performed in Step 2.
The obtained results are depicted in Fig. \ref{Fig_U_k}.

Regarding the range of parameters in \ref{Fig_U_k}, let us note the following. The minimum of both parameters (2) is defined by the assumption that we are interested in non-exclusive (or many-to-many) matchings. The maximum value of $q_T$ is defined by the assumption described in \cite{csercsik2023preallocation}, namely 'every tenant may have at most 8 preallocated channels', which is due to the computational complexity of the CA algorithm following the preallocation process. If more channels would be allocated to any tenant, the computational requirements of the consecutive CA would grow significantly. Regarding the parameter $q_{ch}$, higher values (resulting) may be analyzed also, but as we will see later, they are irrelevant in the context of optimal parameter sets.

Looking at the results depicted in Fig \ref{Fig_U_k}, one may observe that the average total utility value is increased as the size of the scenario increases. This is due to the increasing number of tenants (6/12/20 in the case of SS/MS and LS respectively). We also have to note that the scenarios were defined in order to provide challenging allocation tasks for the algorithms (utility values are bound from above by 6/12/20  respectively), to be able to compare the efficiency of different methods and parameter sets in non-trivial tasks (if there is no resource scarcity, it is not a challenge to provide a good allocation).

Regarding the efficiency of parametrization, on the one hand we can see that if $q_T$ is to small, thus only a few channels are preallocated to tenants, the search space of the CA in the second step is too much constrained to reach an efficient solution (i.e. the tenants do not have enough alternatives). This is quite straightforward, and not surprising. 

On the other hand however, we can see that as $q_{ch}$ is increased, the results tend to deteriorate, which is a more interesting phenomenon. The explanation for this is the following:
Even if we choose $q_T=8$, tenants with free slots (in extreme cases even with no allocated channels) may remain after the execution of the M2MGS -- the probability of such cases and the extreme cases of 'preallocation starvation' (i.e. the event, when no channels are preallocated to a tenant) depend on the number and quota ($q_{ch}$) of channels.
As $q_{ch}$ is increased, single channels in preferred positions (belonging to BSs, which provide good connection to multiple tenants) are able to occupy a higher number of tenant slots (as they provide more offers). In such cases channels of less preferred BS are more likely to be not preallocated at all, implying unused resources.

\begin{figure}
  \centering
  \includegraphics[width=0.48\linewidth]{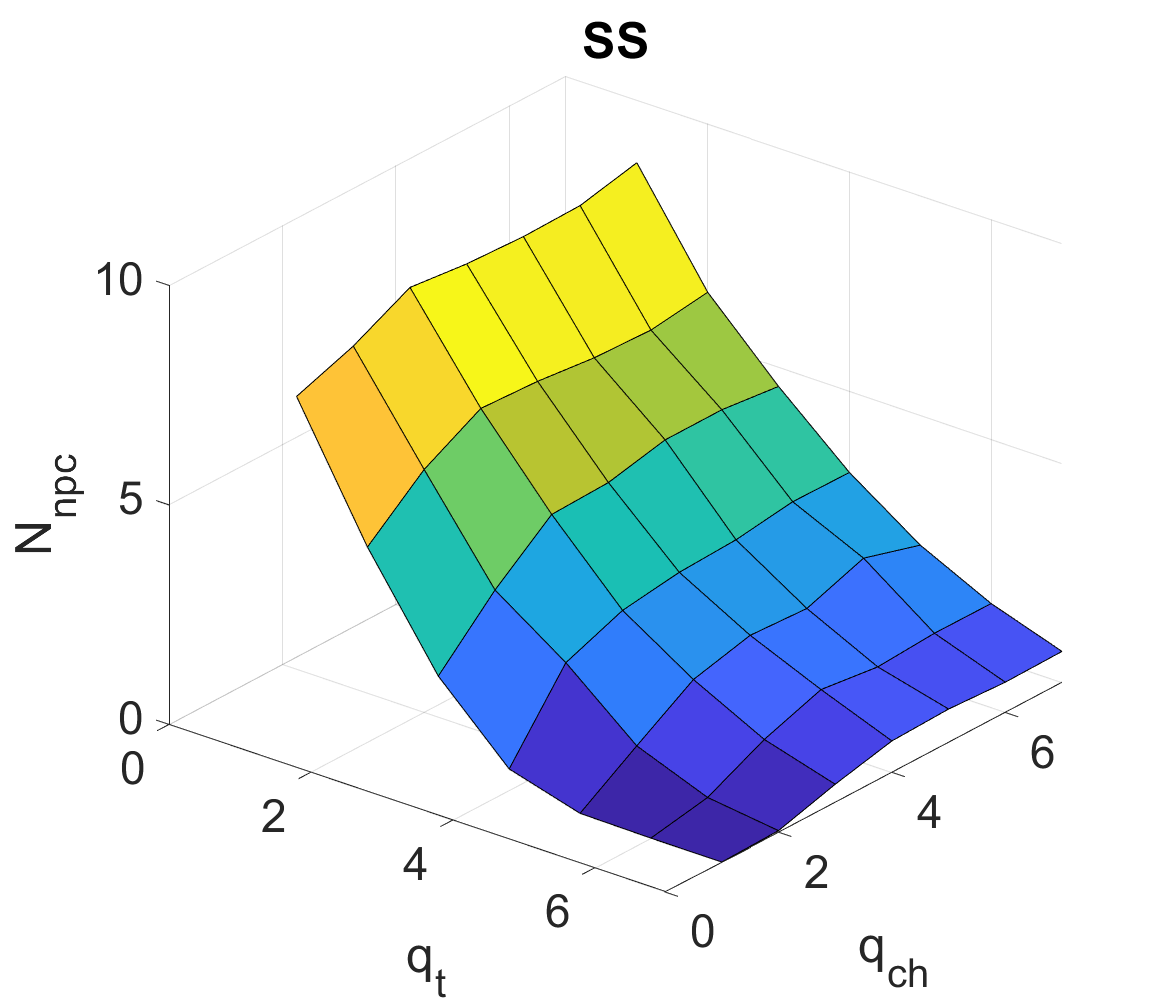}
  \includegraphics[width=0.48\linewidth]{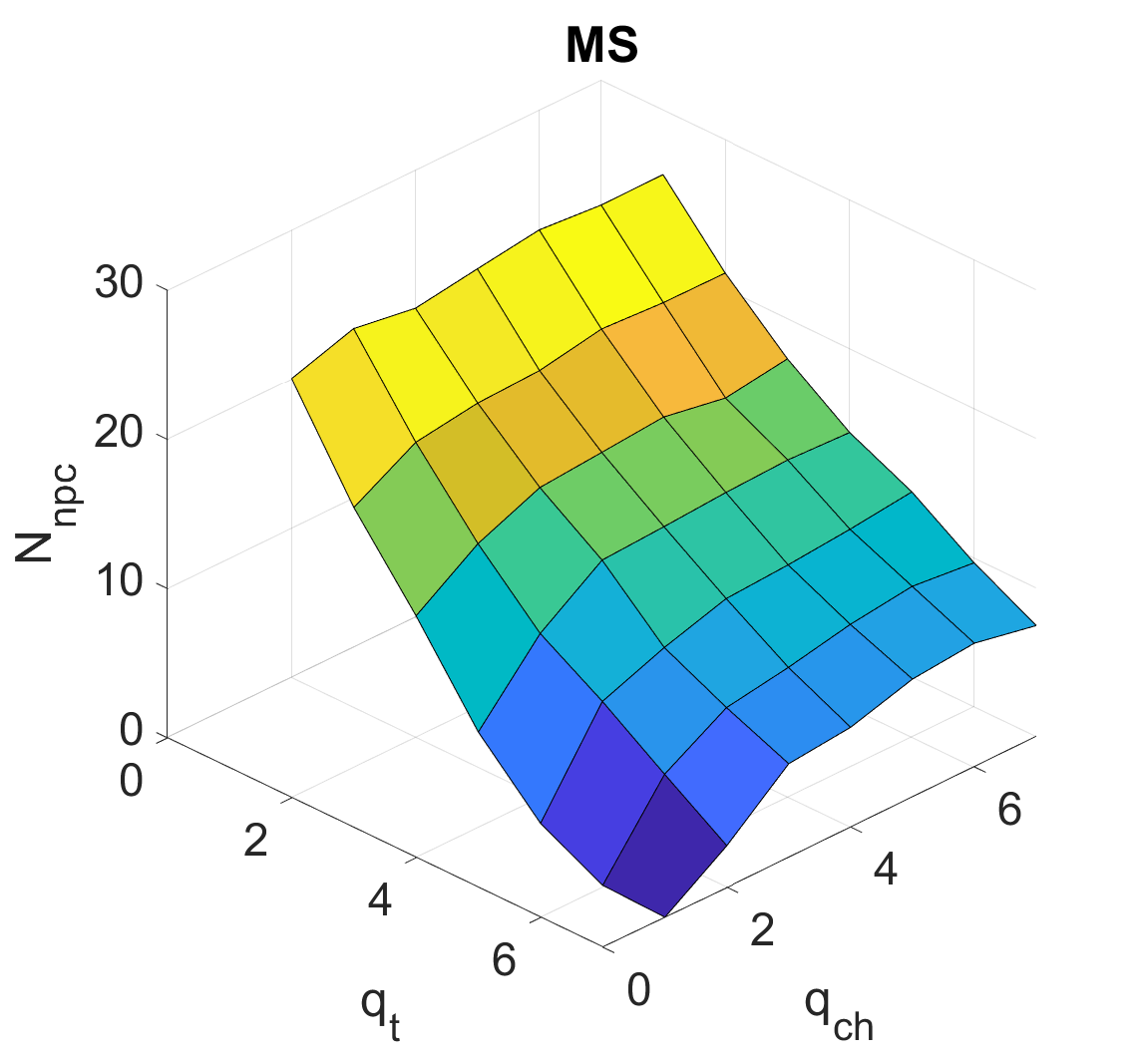}
  \\
  \includegraphics[width=0.48\linewidth]{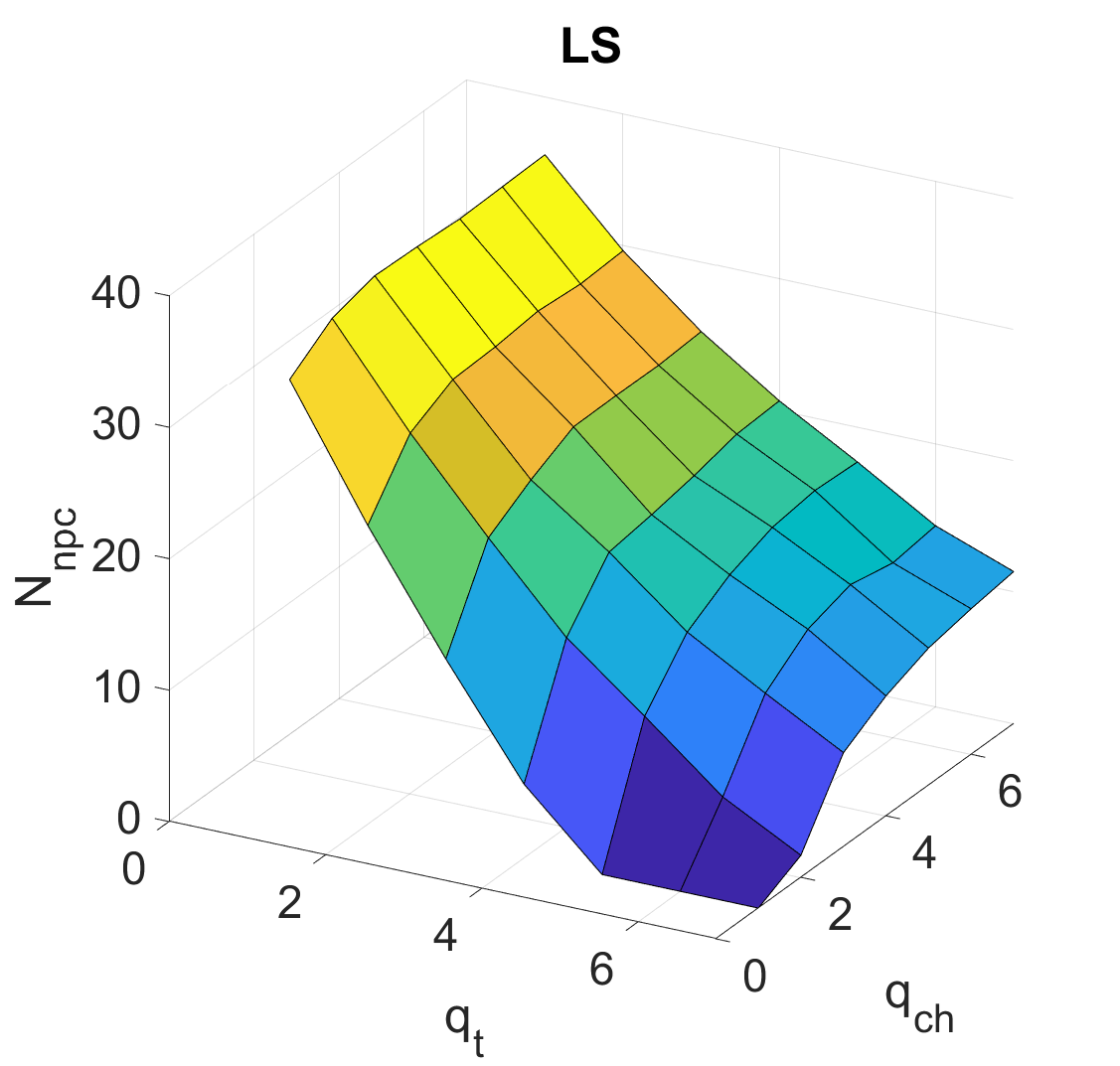}
  \caption{Mean number of not preallocated channels ($N_{npc}$), in the case of various $q_T$ and $q_{ch}$ values. Average values of 200 simulation runs.}\label{Fig_N_npc}
\end{figure}

This is clearly visible in Fig. \ref{Fig_N_npc}, which summarizes the values corresponding to the average number of not preallocated channels (channels which are not assigned to any tenant in the preallocation step) under different values of $q_{ch}$ and $q_T$. Very small and zero values in Fig. \ref{Fig_N_npc} correspond to cases, when it is likely, or sure that the total number of channels ($n^T_{ch} \times q_{ch}$) is less or equal compared to the total number of tenants ($n_T \times q_{T}$), thus all channels are preallocated (the number of channels is randomized with upper and lower bounds, see the description of setups in subsection \ref{sec_params}).

\begin{figure}
  \centering
  \includegraphics[width=0.6\linewidth]{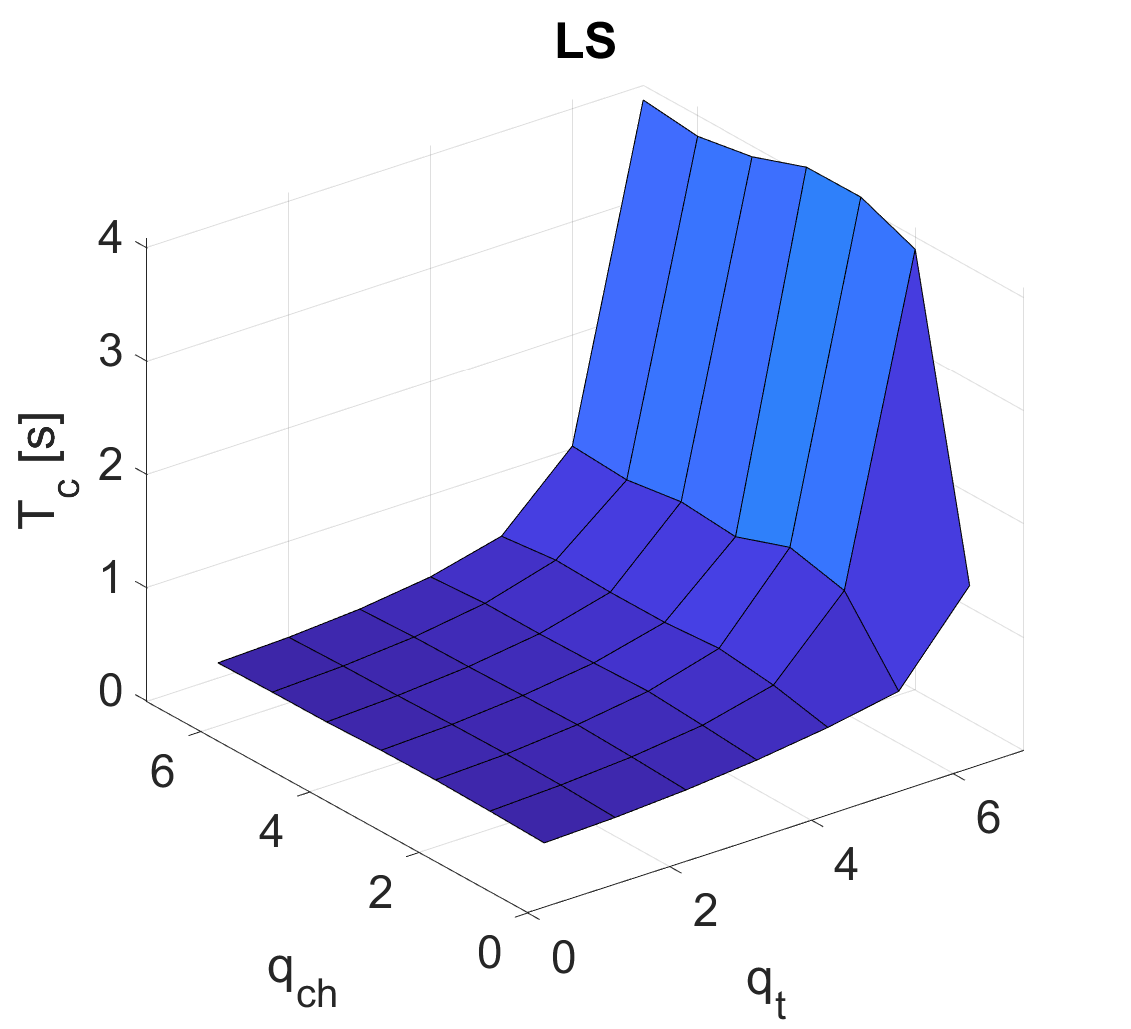}
  \caption{Average value of computational time required for the preallocation process [s], for various $q_T$ and $q_{ch}$ values in the case of the LS setup (200 simulation runs).}\label{Fig_T_c_LS}
\end{figure}

One additional important factor is the computational time required for the preallocation process. Fig. \ref{Fig_T_c_LS} depicts the respective results for the LS setup, which show that the required computational time increases significantly as $q_T$ is increased (channels propose). Regarding scalability, we can say that while the trends regarding the dependence on $q_t$ and $q_{ch}$ are very similar in the case of the SS and MS setups, the value of the required computational time is decreased by 90\% and 70\% respectively in these cases.
The simulations were run on a standard desktop PC (Intel core i5 @ 2.9 GHz, 16 GB RAM, 64 bit OS, MATLAB environment).

\subsection{Discussion}

Overall, we can observe that in the context of utility values, the most efficient preallocations are resulting from cases when $q_T$ is high but $q_{ch}$ is low. At first glance, this may be surprising, since if a channel is e.g. only preallocated to two tenants, one may assume that the CA in the second step will have a narrow search space. However, if we have e.g. 15 channels (a realistic number in the SS setup), this still means $2^{15}$ possible cases. In addition, the phenomenon discussed earlier in the case of not preallocated channels may be also relevant. If a channel has high quota, and if it is popular (i.e. it belongs to a BS with good connection parameters to multiple tenants), it will be preallocated to the maximal $q_{ch}$ number of tenants, although in the final allocation, it may be assigned to only one of them. Nevertheless, it occupies multiple slots of tenants, and thus supersedes other potential channels as alternatives for the respective tenants.

\section{Conclusions}

In this paper we have shown that the appropriate choice of parameters is critical in the case of the M2MGS preallocation method. As the results demonstrated, suboptimal parameter sets can easily result in poor performance and/or increased computational time. Inappropriate parametrization may decrease the performance in the terms of the resulting utility by even 50 \%, and/or increase the computational time by appr. 30 \%. Appropriate parameter sets in the case of the M2MGS preallocation method correspond to configurations, when (1) the number of tenant and channel/BS slots allow the preallocation of all or nearly all channels (2) and channels/BSs are preallocated to a low number of tenants ($q_{ch}$ / $q_{BS}$).



\bibliographystyle{IEEEtran}
\bibliography{GS_RR.bib}


\end{document}